# MODELS OF SOFT X-RAY TRANSIENTS AND DWARF NOVAE

J.P. LASOTA

*UPR 176 du CNRS; DARC*
*Observatoire de Paris, Section de Meudon*
*92190 Meudon Cedex, France*

**Abstract.** Models of Soft X–ray Transients are presented and compared with observations. The importance of inner advection–dominated flows in quiescent transient sources is discussed, as well as the problem of global stability of the standard outer accretion disc. A comparison is made with similar problems in dwarf nova models.

## 1. Introduction

Soft X–ray Transients (SXT) are low–mass X–ray binaries (LMXB) which spend most of the time in a very dim and quiescent state, but on rare occasions undergo outbursts during which they can become the brightests X–ray sources on the sky. Like other LMXBs, SXTs are close binaries in which a compact object, a black hole or a neutron star, accretes matter from a low mass companion: a lower main sequence star or a (sub)giant. The risetimes to outburst in SXTs are typically 2 to 10 days, the main outburst lasts from 20 to 90 days; but in some cases weaker outbursts can be observed for the year after the main event. The luminosity at outburst rises to $10^{37} - 10^{38}$ erg s$^{-1}$ and the total energy released is $\sim 10^{43} - 10^{44}$ erg. It seems that the recurrence time for SXTs containing black holes is longer than in systems containing neutron stars.

The name 'Soft X-ray Transient' derives from the fact that most of the outburst energy is emitted in rather soft (few keV) X-rays. It has been claimed that spectra of black–hole SXTs are characterised by an 'ultra–soft' component, but in a few cases, in which the accreting object is most probably a black hole, such a component has not been observed (Harmon et al. 1994). Black hole SXTs emit also very high energy photons, in the case of GRS 1124-684 an $e^+$–$e^-$ annihilation line was observed (Goldwurm et al. 1993). On the other hand neutron star SXTs were observed to emit above 100 keV (Barret et al. 1993).

Black hole SXTs (BSXT) are sometimes called 'X–ray novae'. This is confusing because, the name 'Nova Muscae', for example, is used to describe both the BSXT GRS 1124-684 and the classical nova GQ Mus, which was also an X–ray source.

The mechanism responsible for SXT outbursts is unknown. It is clear that the site of the eruption is the accretion disc, but it is still a subject of debate whether it is due to an instability in the disc or to an enhanced mass transfer from the companion, or to a combination of the two (see e.g. Lasota 1995). There is an obvious analogy between SXTs and dwarf novae (van Paradijs & Verbunt 1984) in which the black hole or the neutron star is replaced by a white dwarf. In particular dwarf nova superoutbursts, which have bigger amplitudes and last longer than the 'normal' (U G Gem – type) outbursts are, in many respects similar to SXT outbursts.

In the case of of dwarf novae (DN) the disc instability model (DIM) seems to account for the most general properties of 'normal' outbursts (Cannizzo 1993), even if it requires some modifications and additions (Ludwig et al. 1994; Livio & Pringle 1992; Meyer & Meyer-Hofmeister 1994) which concern the inner disc structure.

In the case of superoutbursts, two kind of models have been considered. On the one hand Osaki (1989) has proposed a model in which superoutbursts result from a combination of thermal and tidal instabilities. On the other hand Osaki (1985), Whitehurst & King (1991), Whitehurst (1994) and Smak (1995) have suggested that an enhanced mass transfer (EMT) from the companion is an important factor in the superoutburst mechanism.

As for DN, two kinds of models have been proposed to explain the SXT phenomenon: the DIM and the EMT model. The DIM (Huang & Wheeler 1989; Mineshige & Wheeler 1989; Cannizzo et al. 1995) was an extension of the DN DIM to systems containing black holes (no model for neutron star SXTs has been proposed). Hameury, King & Lasota (1986, 1997, 1988, 1990) worked on a model in which EMT was triggered by X-ray illumination of the secondary during quiescence. Both models had difficulties with accounting for the observed SXT properties (Lasota 1995) but the X-ray EMT model can be ruled out for systems with orbital periods $P_{\rm orb} \lesssim 10$ h (Gontikakis & Hameury 1993; Lasota 1995). The DIM for SXTs is contradicted by X–ray observations (Mineshige et al. 1992; Verbunt et al. 1994; Wagner et al. 1994; McClintock, Horne & Remillard 1995; Lasota 1995; Verbunt 1995) of transient systems in quiescence.

## 2. The standard disc model of SXTs

The DIM requires the whole disc in quiescence to be in a cold 'low state'; or in another words, it must be everywhere (at each radius) on the lower branch of the 'S–curve' representing the local equilibria of the accretion disc. This requirement means that the surface density of the quiescent disc $\Sigma$ has to be less than the critical surface density $\Sigma_{\max}(r)$ everywhere in the disc (see Fig. 1). This implies that an inequality has to be satisfied by the accretion rate in a quiescent disc (Lasota 1995):

$$\dot{M}_{\rm in} \lesssim \frac{8\pi}{3}\nu\Sigma_{\max} \approx 2.76 \times 10^4 t_8^{-1} r_7^{3.11} m_{10}^{-0.37} \alpha^{-0.79} \quad {\rm g\ s}^{-1} \qquad (1)$$

where $\nu$ is the kinematic viscosity coefficient, $m_{10}$ is the mass of the compact object in units of $10{\rm M}_\odot$, $t_8$ is the recurrence time in units of $10^8$ s, $r_7$ is the radius in $10^7$ cm and $\alpha$ is the Shakura–Sunyaev viscosity parameter. Numerical models by Mineshige & Wheeler (1989; MW) satisfy the inequality given by Eq. (1).

*Figure 1.* Surface density profiles of a truncated accretion disc around a black hole with mass $M_1 = 4.4{\rm M}_\odot$, for five values of accretion rates: $10^{13}, 10^{14}, 10^{15}, 10^{16}, 10^{17}$ g s$^{-1}$. $r_9$ is the radius in $10^9$ cm. For a given $\dot{M} = const$ the disc can be in a globally stable *cold* equilibrium only for radii larger than the radius at which the density profile becomes tangent to the $\Sigma_{\max}$ line.

Several black-hole SXTs have been observed by GINGA (Mineshige et al. 1992) and ROSAT (see Verbunt 1995). Two systems, A0620-00 and V404 Cyg, have been detected at levels corresponding to accretion rates of at least $\sim 1.4 \times 10^{10}$ g s$^{-1}$ for A0620-00 and of $3 \times 10^{12}$ g s$^{-1}$ for V404

Cyg. Those accretion rates are several orders of magnitude larger than the values predicted by the disc instability model of MW. Detections of X-ray radiation from quiescent SXTs contradict the disc instability model, at least in the version proposed by MW (Mineshige et al. 1992).

In the case of A0620–00, the prototypical BSXT, the optical luminosity suggests a mass accretion rate of $\sim 6 \times 10^{15}$ g s$^{-1}$ in the outer part of the accretion disc (McClintock et al. 1995), five orders of magnitude higher than the accretion rate at the inner disc as deduced from observations (and 10 orders of magnitude higher if one uses Eq. (1) as required by DIM!).

The accretion rate on to the compact object as given by X–ray observations of quiescent SXTs poses an insurmountable obstacle to models in which the inner accretion disc is of the Shakura–Sunyaev type. On the one hand the requirement of Eq. (1) can agree with observations only if one assumes a very low viscosity (e.g. $\alpha \sim 10^{-8}$), in fact no viscosity at all, in the inner disc. On the other hand X–rays have to be produced by a process connected to viscosity if one assumes a standard disc structure.

## 3. Advection–dominated quiescent SXT accretion discs

Since X–rays in quiescent SXTs are produced by accretion and not emitted by the companion (Verbunt 1995), one should consider other accretion flow structures that could describe quiescent SXTs.

Narayan et al. (1995) have proposed a model of quiescent BSXTs in which the inner ($r \lesssim 10^3 r_S$, where $r_S = 2GM/c^2$) accretion flow is advection–dominated. The standard, Shakura-Sunyaev disc model assumes that the energy released through viscous dissipation is radiated locally and efficiently. In such a disc the efficiency of converting gravitational potential energy into luminosity is close to $r_{\rm in}/r_S$, which for neutron stars and black holes is $\sim 0.1$. In the case of inefficient cooling, for example when the accretion flow is optically thin, another mode of cooling may dominate: the viscously released energy is advected with the gas rather than being radiated. Such "advection–dominated" flows have been studied first in the optically thick (radiation pressure dominated) case (Begelman & Meier 1982; Abramowicz et al. 1988) and more recently for the optically thin case (Abramowicz et al. 1995; Chen et al. 1995; Narayan & Yi 1994, 1995ab, Abramowicz & Lasota 1995).

The main difference between the description of 'standard' geometrically thin accretion disc and advection–dominated accretion flows is in the energy equation

$$Q_{\rm adv} \equiv \Sigma v_r T \frac{ds}{dr} = Q^+ - Q^- \qquad (2)$$

Figure 2. Equilibria of accretion flows on to a black hole in the surface density – accretion rate plane for a central mass $10 M_\odot$, and $\alpha = 0.01$ at a distance of $10 r_S$. The advection–dominated solutions are indicated by "A", the radiatively cooled, stable solutions by "R" and the radiatively cooled, unstable solutions by "U". Viscously stable branches have positive slopes, thermally stable branches have $Q^+ > Q^-$ below and $Q^+ < Q^-$ above them. The total cooling rate $Q^-$ includes all radiative processes *and* advection.

where $v_r$ is the radial speed, $T$ is the temperature and $s$ the entropy. In the standard model of thin accretion disks (Frank et al. 1992) advective cooling is neglected because the ratio of advective to cooling $Q^{\rm adv}$ to the viscous heating $Q^+$ is proportional to the square of the ratio of the disc thickness $H = H(r)$ to the radius $r$,

$$Q^{\rm adv} \sim \left(\frac{H}{r}\right)^2 Q^+ \qquad (3)$$

and, of course, for thin disks $H \ll r$. Advection-dominated flows are much hotter than cooling-dominated thin disks and are therefore thicker in the vertical direction. For this reason they are also described as "slim discs" or "thick discs" depending on the relative vertical thickness.

Figure 2. shows an example (Abramowicz et al. 1995; Abramowicz & Lasota 1995) of accretion flow equilibria with the energy equation in the form of Eq. (2). In this case the radiative cooling term contains only bremsstrahlung but the shape and form of the solutions represented here is very general (Chen et al. 1995). The $S$–shaped line corresponds to optically thick solutions, "slim discs" (Abramowicz et al. 1988). The curve in the lower left corner represents optically thin solutions of which those marked

by "U" are radiatively cooled and thermally unstable. The advection–dominated solutions indicated by "A" are both viscously and thermally stable so they represent the only physically possible hot, optically thin accretion flow configurations.'

In an advection–dominated flow most of the energy that is realeased by viscous heating is retained in the gas and swallowed by the black hole. Only a very small fraction of this energy is radiated away so that the radiative efficiency is $\ll 0.1$. In this case one cannot deduce the accretion rate from the formula $\dot{M} \sim 10Lc^2$.

*Figure 3.* The spectrum of the quiescent BSXT A0620-00 according to Narayan et al. (1995). The filled circles show measured fluxes of A0620-00 plotted as $\nu F_\nu$ vs $\nu$. The ROSAT point at $\nu \sim 10^{17}$ Hz is shown along with a "bow-tie" to indicate the range of spectral slopes allowed by the data. The downward pointing arrows represent the various upper limits in radio, infrared and EUV. The lines show the spectrum predicted by the model. The parameters of the model are indicated on the left, where $\alpha = 0.1$ is the viscosity parameter in the advection–dominated flow, $\beta$ is the ratio of the gas to the total (gas plus magnetic) pressure, $i$ is the inclination, $M$ the black hole mass and $\dot{M}$ the (constant through the flow) accretion rate. $v_{\rm in}$ and $v_{\rm out}$ correspond respectively to the inner and outer radii of the 'standard' thin disc. $D$ is the distance to the system The dotted line shows the contribution to the spectrum from the outer thin disc, the dashed line corresponds to the inner hot flow, and the solid line is the combined spectrum.

Narayan et al. (1995) have presented models of the BSXTs A0620-00, V404 Cyg and GRS 1124-684 in quiescence in which the accretion flow consists of two parts: an outer standard thin disc, extending outward from $\sim 3000 r_S$, and an inner, two-temperature, advection-dominated flow described by the Narayan & Yi (1995b) solution in which cooling by Comptonized synchrotron and bremsstrahlung is included. The result for A0620-00 is shown on Fig. 2. The model is consistent with all the available data. In the hot, advection–dominated flow only a a very small fraction ($10^{-3}-10^{-4}$) of the thermal energy is radiated away in the inner flow regions. In this model the flow is stationary; the accretion rate ($\dot{M} = 4 \times 10^{14}$ g s$^{-1}$ for A0620-00) is constant everywhere.

## 4. Equilibria of standard thin discs at very low accretion rates

In their model Narayan et al. (1995) assume that the outer standard disc is stationary. This disc extends outwards from a radius $\sim 10^3 r_S$, which for the compact object masses of interest is $\sim 10^9$ cm. Since the accretion rates considered are $\lesssim 10^{15}$ g s$^{-1}$, the outer disc can be described as an accretion disc in a cataclysmic variable where the radius of the compact object (a white dwarf) is $\sim 10^9$ cm. Equilibria of accretion discs in CVs were studied in detail in the DIM for dwarf novae (see Cannizzo 1993 for a review).

At a given radius $r$, thermal equilibria of accretion discs in cataclysmic variables form, on the $\Sigma$–$\dot{M}$ (or $\Sigma$–$T_{\mathrm{eff}}$) plane, an S–shaped curve, where the middle branch with the negative slope is thermally (and viscously) unstable. Hot stable equilibria therefore exist only for $\Sigma > \Sigma_{\min}$ and cold ones only for $\Sigma < \Sigma_{\max}$ (see e.g. Cannizzo 1993). For the disc to be globally unstable, i.e. unstable at some radius $r_{\mathrm{in}} \leq r \leq r_{\mathrm{out}}$, the mass transfer rate $\dot{M}_T$ must satisfy the inequality

$$\dot{M}\left(\Sigma_{\max}(r_{\mathrm{in}})\right) < \dot{M}_T < \dot{M}\left(\Sigma_{\min}(r_{\mathrm{out}})\right) \tag{3}$$

The range of unstable mass transfer rates therefore depends on the size of the disc. Fig. 1 shows surface density profiles of (truncated) equilibrium accretion discs around a 4.4M$_\odot$ black hole (the value in the Narayan et al. (1995) model of of A0620). The value of the viscosity parameter $\alpha$ is assumed to be 0.01. One should note that there is no reason for $\alpha$ to be the same in the advection–dominated and standard regions; on the contrary, one should expect the viscosity to be higher in the hot flow. The accretion rate $\dot{M}$ is constant for each curve. Models by Hameury et al. (1995) were used in this calculation.

The part of the curve which is above the $\Sigma_{\max}$ line cannot represent a cold thermal equilibrium so that the disc cannot be globally in equilibrium ($\dot{M} = const$) for the corresponding mass transfer rate.

From Fig. 1 one can see that there exist two types of equilibria represented by the $\Sigma(r)$ curves: for $\dot{M} \lesssim \dot{M}'$ (where $\dot{M}'$ is some value of accretion rate depending on $M_1$ and $\alpha$; in Fig. 1 $\dot{M}' \sim 10^{13} - 10^{14}$ g s$^{-1}$) the surface density is always less than $\Sigma_{\max}$. For such low $\dot{M}$ the disc is globally in cold stable equilibrium, from $r_{\mathrm{out}}$ down to $r_{\mathrm{in}}$. For $\dot{M} > \dot{M}'$ only the outer part of the disc can be in a cold stable equilibrium. Starting at some $r_{\rm out}$ from $\Sigma < \Sigma_{\rm max}$, the $\Sigma(r)$ curve reaches, with decreasing $r$, the value $\Sigma_{\rm max}$ at some radius $r_{\rm crit}$. The disc can be in cold equilibrium only for $r_{\rm crit} < r < r_{\rm out}$. The next segment of the curve between $\Sigma_{\rm max}$ and $\Sigma_{\rm min}$ is unstable while the curve to the left of $\Sigma_{\rm min}$ represents hot equilibria. For $\dot{M} > \dot{M}'$ the disc is therefore globally unstable down to $r_{\rm in}$, and a cold steady state cannot exist.

There exists however a radius $r_{\rm in} = r_{\rm crit}$ for which the disc is globally stable for some $\dot{M} = const$. From the example shown on Fig. 1. one can deduce that the global state of the disc will depend on the exact value of $r_{\rm in}$. If one takes the 'best fit' values shown of Fig. 3. in the model of Narayan et al. (1995), i.e. $M_1 = 4.4 {\rm M}_\odot$, and $r_{\rm in} = 1800 r_S$ (corresponding to $v_{\rm in} = 5000$ km s$^{-1}$ and $\dot{M} = 4 \times 10^{14}$ g s$^{-1}$), the disc will be globally unstable in quiescence. For $r_{\rm in} = 5000 r_S$, however, the disc would be globally stable. The uncertainty in the system's (see e.g. Shahbaz et al. 1994) and in model parameters does not allow a definitive conclusion about the global stability of quiescent SXT outer discs.

A similar problem has been studied by Lasota et al. (1995) in the context of dwarf nova outbursts. Some SU UMa–type dwarf novae show only very rare and very long superoutbursts and no (or almost no) 'normal' outbursts. The best known system in this class is WZ Sge. On the other hand, DIM applied to dwarf novae requires some modifications to be applicable to the whole class of observed events. In particular the so–called UV–delay problem and the X–ray and UV emissions from quiescent DNs require a 'hole' in the inner disc regions (see e.g. Meyer & Meyer–Hofmeister 1994). This 'hole' can be produced by magnetic disruption if the white dwarf is (weakly) magnetized (Livio & Pringle 1992) or it can be the result of evaporation (Meyer & Meyer–Hofmeister 1994) of the flow which is unable to radiate the viscously released energy. The second mechanism was proposed by Narayan & Yi (1995) to explain the origin of advection–dominated flows around black holes (see below).

Lasota et al. (1995) show that if the inner regions of accretion discs in quiescent dwarf nova systems are removed, the remaining disc is globally stable for mass transfer rates of $\lesssim 10^{15}$ g s$^{-1}$. This implies that (super)outbursts in these systems have to be triggered by an enhanced mass transfer from the companion. They suggest that the lack of normal outbursts in WZ Sge results only from its low mass transfer rate: there are no outbursts because the disc is stable. A superoutburst would be triggered by an EMT which put the disc into a globally unstable state; in other words, superoutbursts would be due to a disc instability generated by an increased mass transfer. There are observational reasons to think that such a hybrid mechanism is at work in SU UMa's (Smak 1995). The viscosity in WZ Sge disc does not therefore have to be a few orders of magnitude lower than in other quiescent dwarf novae.

## 5. Soft X–ray transient outbursts

Fig. 2. shows that (at a given radius) for a whole range of accretion rates no stable equilibria can exist. In the example shown, above $\sim 0.03 \dot{M}_{\rm E}$ no optically thin solution exists, and above $\sim 0.3 \dot{M}_{\rm E}$ no stable configuration is possible. An advection–dominated solution for a quiescent SXT will correspond to a point at the lower branch marked by an "A". If for some reason (increased mass transfer rate or outer disc instability) the mass accretion rate in the inner flow is increased to a value for which no optically thin solution exists, the system will become unstable. The system point will have to move in the $\Sigma$–$\dot{M}$ plane but the whole process will be non–local and complicated by the fact that $r_{\rm in}$ will (probably) move inwards and the outer disc will invade the inner hot flow (see Narayan et al. 1995).

As mentioned above the outer disc may be stable or unstable, depending on the value of its inner radius. In the first case (Yi et al. 1995) SXT outbursts would be generated by an EMT from the companion, as in the scheme proposed by Lasota et al. (1995) for WZ Sge: SXTs would be WZ Sge-type systems with a black hole instead of a white dwarf. The increased mass transfer is in this case only a 'detonator': it makes the disc unstable so that the outburst is finally due to a disc instability.

In the second case (Mineshige 1995, Yi et al. 1995) the outer disc is unstable and undergoes dwarf nova type instabilities. In this case, however, contrary to the case of WZ Sge (Smak 1991), no abnormally low viscosity is required to get the right recurrence time etc. (Yi et al. 1995).

I am grateful to Jean-Marc Huré for producing Figure 1. I thank Didier Barret, John Cannizzo, Jeff McClintock, Jean-Marie Hameury, Emmi Meyer–Hofmeister, Andrew King, Friedrich Meyer, Ramesh Narayan, Joe Patterson, Rob Robinson, Joe Smak and Insu Yi for interesting and stimulating discussions.

## References

Abramowicz, M.A. & Lasota, J.P. (1995) *Comments in Astrophys.*, in press
Abramowicz, M.A., Czerny, B., Lasota, J.P. & Szuszkiewicz, E. (1988) *Astrophys. J.* **332**, 646
Abramowicz M.A., Chen, X.M., Kato, S., Lasota, J.P. & Regev O. (1995) *Astrophys. J.* **438**, L37
Barret, D. et al. (1993) *Astrophys. J. Sup. Ser.* **97**, 241
Begelman, M.C. & Meier, D.L. (1982) *Astrophys. J.* **253**, 873
Cannizzo, J.K. (1993) in *Accretion Disks in Compact Stellar Systems*, ed. J. Craig Wheeler, World Scientific Publishing, Singapore, p.6
Cannizzo, J.K., Chen, W. & Livio, M. (1995) *Astrophys. J.*, in press

Chen, X.-M., Abramowicz, M.A., Lasota, J.P., Narayan, R. & Yi. I. (1995) *Astrophys. J. Lett.* **4**, 43, L61
Frank, J., King, A.R. & Raine, D. (1992) *Accretion Power in Astrophysics*, CUP: Cambridge
Goldwurm et al. (1993) *Astrophys. J. Sup. Ser.* **97**, 293
Gontikakis, C. and Hameury, J.M. (1993) *Astron. Astrophys.* **271**, 118
Hameury, J.M., King, A.R. and Lasota, J.P. (1986) *Astron. Astrophys.* **162**, 71
Hameury, J.M., King, A.R. and Lasota, J.P. (1987) *Astron. Astrophys.* **171**, 140
Hameury, J.M., King, A.R. and Lasota, J.P. (1988) *Astron. Astrophys.* **192**, 187
Hameury, J.M., King, A.R. and Lasota, J.P. (1990) *Astrophys. J.* **353**, 585
Hameury, J.M., Huré, J.M. & Lasota, J.P. (1995) *Astron. Astrophys.*, in preparation
Harmon, B.A. et al. (1994) in *The Second Compton Symposium*, ed. C. Fichtel. N. Gehrels & J.P. Norris, New York: AIP, p.210
Huang, M. and Wheeler, J.C. (1989) *Astrophys. J.* **343**, 229
Lasota J.P., (1995) in *Compact Stars in Binaries*, Proceedings of the IAU Symposium 165, eds. van den Heuvel E.P.J, van Paradijs J., Kluwer, in press
Lasota, J.P., Hameury, J.M. & Huré, J.M. (1995) *Astron. Astrophys.*, submitted
Livio M., Pringle J.E., (1992) *Mon. Not. R. astr. Soc.* **259**, 23P
Ludwig K., Meyer-Hofmeister E. & Ritter H. (1994) *Astron. Astrophys.* **290**, 473
McClintock, J.E., Horne, K. & Remillard, R.A. (1995) *Astrophys. J.*, in press
Meyer F. & Meyer-Hofmeister E. (1994) *Astron. Astrophys.* **288**, 175
Mineshige, S. (1995) *Astrophys. J.*, submitted
Mineshige, S. and Wheeler, J.C. (1989) *Astrophys. J.* **343**, 241
Mineshige, S. et al. (1992) *Publ. Astro. Soc. Japan* **44**, 117
Narayan, R. & Yi, I. (1994) *Astrophys. J.* **428**, L13
Narayan, R. & Yi, I. (1995a) *Astrophys. J.* **444**, 231
Narayan, R. & Yi, I. (1995b) *Astrophys. J.*, in press
Narayan, R., McClintock, J.E. & Yi, I. (1995) *Astrophys. J.*, in press
Osaki, Y. (1985) *Astron. Astrophys.* **144**, 369
Osaki, Y. (1989) *Publ. Astro. Soc. Japan* **41**, 1005
van Paradijs J. and Verbunt, F. (1984) in *High Energy Transients in Astrophysics*, ed. S.E. Woosley, AIP Conf. Proc. No. 115, p. 49
Shahbaz, T., Naylor, T. & Charles, P.A. (1994) *Mon. Not. R. astr. Soc.* **269**, 756
Smak, J.I. (1991) *Acta Astron.* **41**, 269
Smak, J.I. (1995) these proceedings
Verbunt, F. (1995) in *Compact Stars in Binaries*, Proceedings of the IAU Symposium 165, eds. van den Heuvel E.P.J, van Paradijs J., Kluwer, in press
Verbunt, F.et al. (1994) *Astron. Astrophys.* **285**, 903
Wagner et al. (1994) *Astrophys. J. Lett.* **401**, L97
Whitehurst, R. (1994) *Mon. Not. R. astr. Soc.* **266**, 35
Whitehurst, R. & King, A.R. (1991) *Mon. Not. R. astr. Soc.* **249**, 25
Yi, I., Lasota, J.P. & Narayan, R. (1995) in preparation

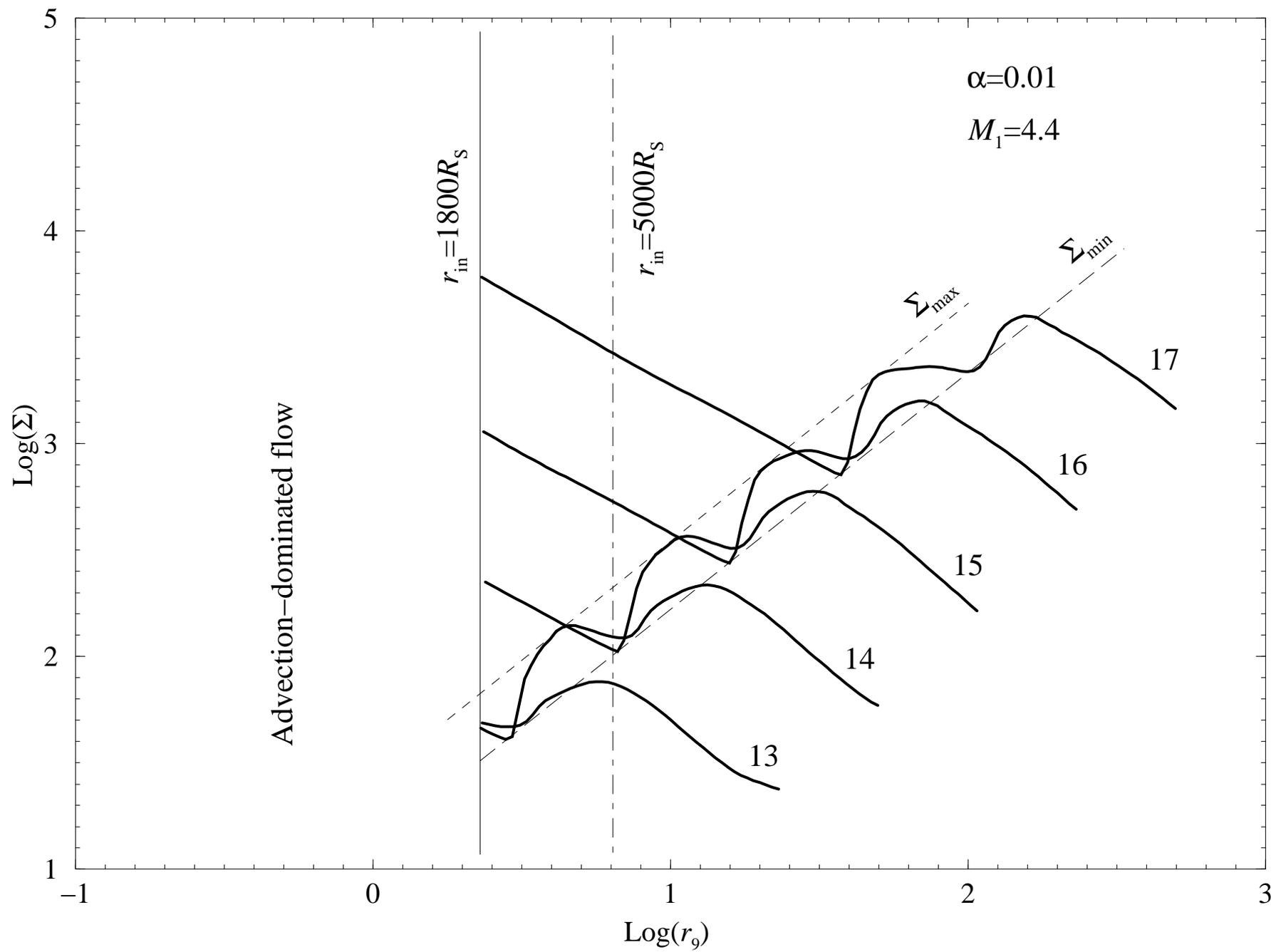

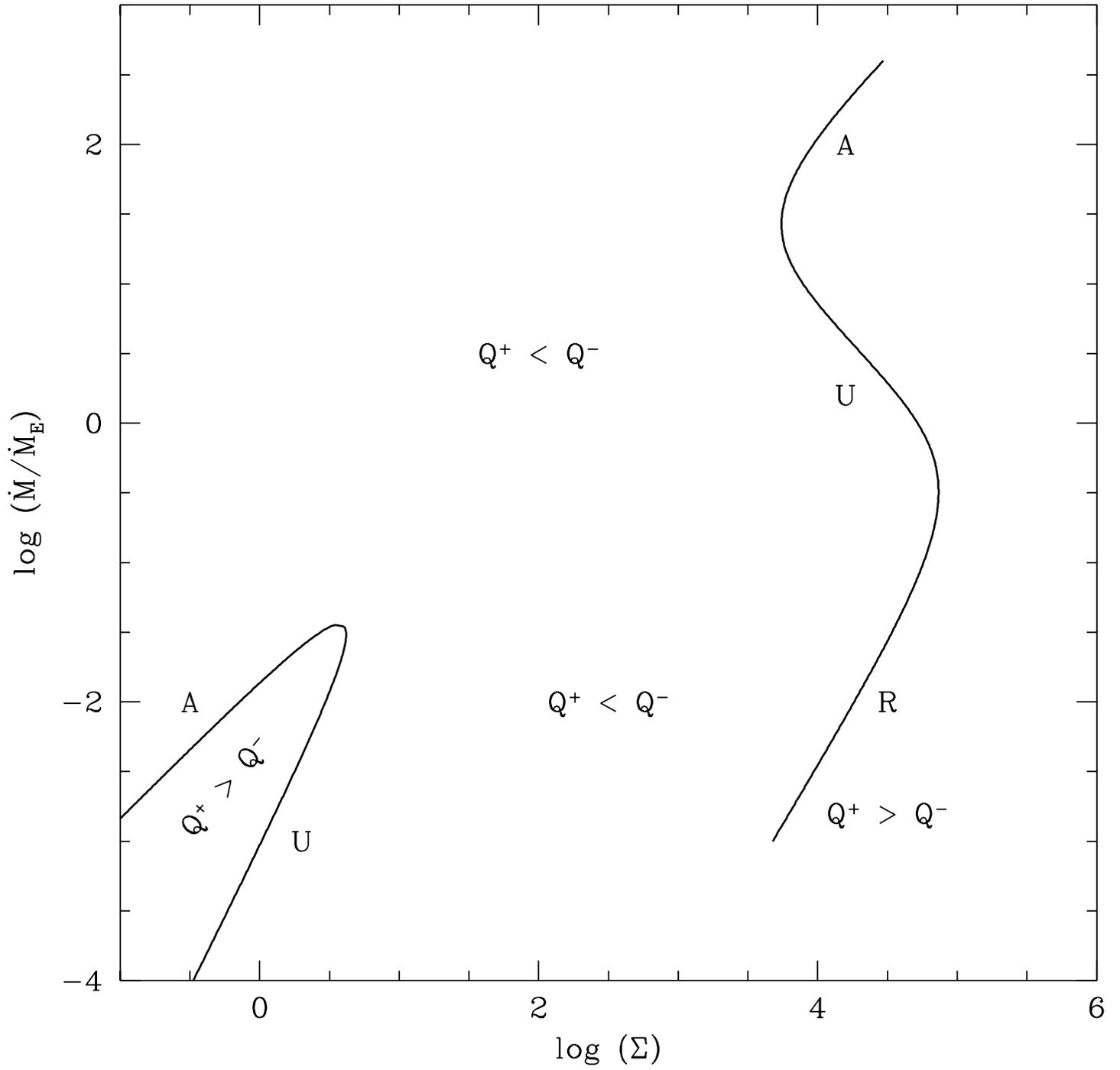

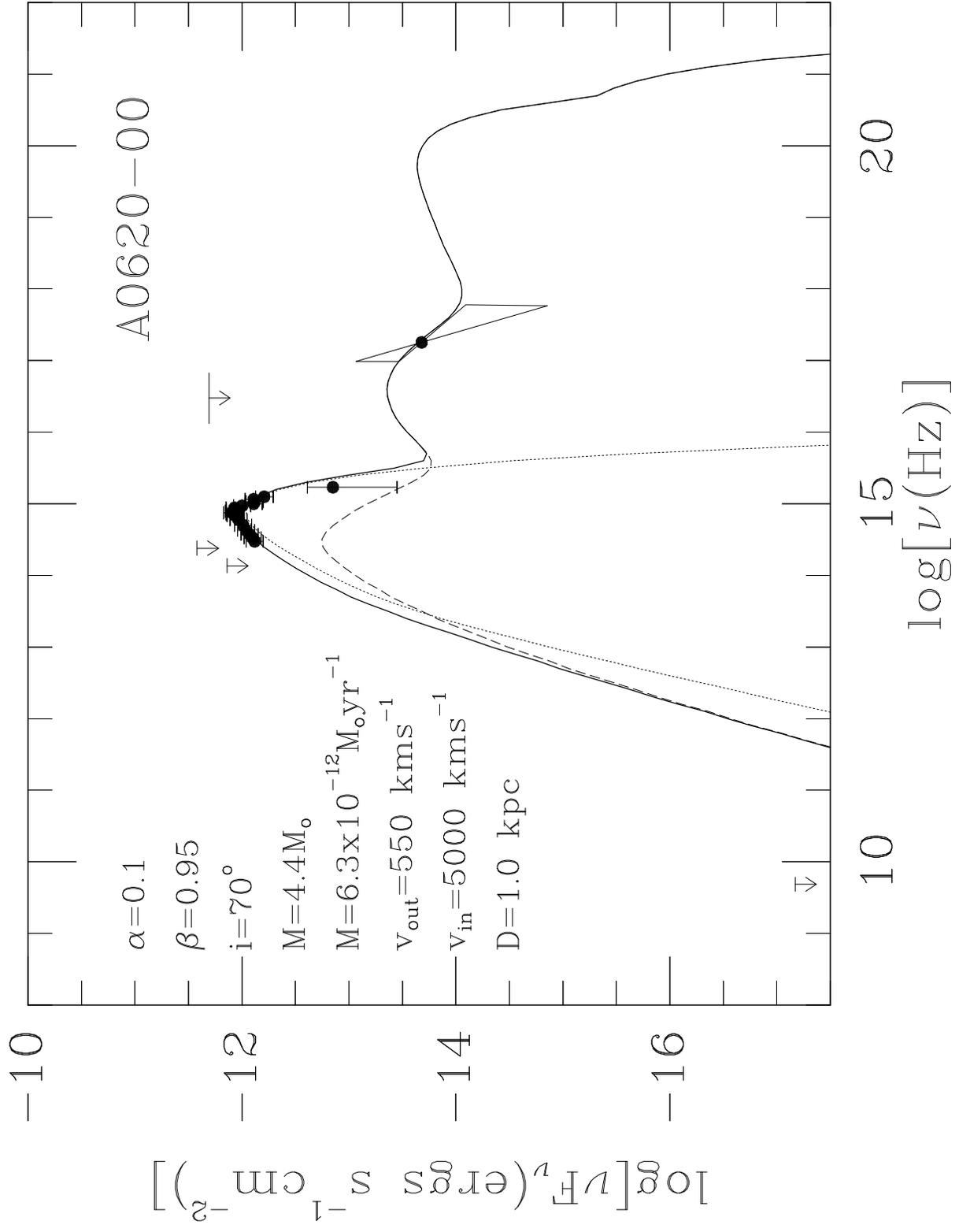